\documentstyle[preprint,aps]{revtex}

\begin{document}
\draft

\preprint{NSF-ITP-95-163}

\title{
A Diagrammatic Theory of Random Scattering Matrices \\
for Normal--Superconducting Mesoscopic Junctions}

\author{N. Argaman and A. Zee}
\address{ Institute for Theoretical Physics,
University of California, Santa Barbara, CA 93106, USA }

\date{Submitted: December 28, 1995}

\maketitle
%\begin{abstract}

\bigskip
\centerline{\large Abstract}
\smallskip
{
\leftskip 0.5in
\rightskip 0.5in

The planar--diagrammatic technique of large--$N$ random
matrices is extended to evaluate averages over the
circular ensemble of unitary matrices.  It is then
applied to study transport through a disordered metallic
``grain'', attached through ideal leads to a normal
electrode and to a superconducting electrode.  The latter
enforces boundary conditions which coherently couple electrons
and holes at the Fermi energy through Andreev scattering.
Consequently, the {\it leading order}\/ of the conductance is 
altered, and thus changes much larger than $e^2/h$ are 
observed when, e.g., a weak magnetic field is applied.  
This is in agreement with existing theories.  The approach 
developed here is intermediate between the theory of dirty 
superconductors (the Usadel equations) and the random--matrix 
approach involving transmission eigenvalues (e.g.\ the DMPK 
equation) in the following sense: even though one starts 
from a scattering formalism, a quantity analogous to the
superconducting order--parameter within the system
naturally arises.  The method can be applied to a variety 
of mesoscopic normal--superconducting structures, 
but for brevity we consider here only the case of a 
simple disordered N--S junction.

}
%\end{abstract}

%73.23.Ps Other electronic properties of mesoscopic systems
%74.50.+r Proximity effects, weak links, tunneling phenomena, 
%                and Josephson effects
% replace with this?  74.80.Fp Point contacts; SN and SNS junctions

\narrowtext

\section{Introduction}

Dissipative single--electron transport in mesoscopic structures
involving normal and superconducting elements has in
recent years become a topic of lively experimental
\cite{Kast,Petr_HM,Petr_Mirr,dV_AI,Po_AI,Petr_AI}
and theoretical
\cite{Nazarov,Zait,Been_RlT,Been_GBP,Brou,p}
activity.  From both of these points of view, this 
is a natural extension of the interest in usual (i.e.,
normal) mesoscopic systems.  Indeed, many of the
technological and mathematical techniques used here are
familiar from that field: lithography, lock--in
amplifiers, diagrammatic perturbation theory, and random 
matrices, to name a few.  However, this extension has often 
lead to surprises.

Consider for example the effect of a magnetic field on a
{\it normal} mesoscopic metallic system, which can be
described in the leading--order approximation by classical
dynamics of non--interacting electrons at the Fermi
surface.  The conductance is then determined by the 
density of states and the geometric details of the sample, 
and is of the order of $N \, e^2/h$ where $N$ is the number 
of propagating modes in the system, $e$ is the electronic 
charge and $h$ is Planck's constant.  In the present work, 
$N$ is taken to be large, whereas additional factors which 
appear in the conductance, such as the factor $l/L$ where 
$l$ is the transport mean--free--path and $L$ is the length 
of the system, are taken to be of order unity and considered 
as fixed geometric characteristics of the system.

If the magnetic field in such a system is so weak that it 
does not affect the classical motion of the electrons, it 
may still shift the phases of different interfering partial
waves in the Schrodinger equation, thus leading to changes
in the conductance which are well--known \cite{wellknown}
to be universal, of the order of the quantum unit of
conductance, $e^2/h$.  This contrasts sharply with the
situation in the presence of one or more superconducting
electrodes (we use ``electrode'' rather than the more
technical nomenclature --- ``particle reservoir'').
By the process of Andreev reflection \cite{Andreev,BTK}, an
electron--like excitation in the normal part of the system 
may reflect off the normal--superconducting (N--S) boundary 
and give rise to a hole--like excitation.  The extra
charge of two electrons is thus transformed into a Cooper
pair, which is carried away by the superconductor.
The Andreev--reflected hole is necessarily found to be in
the state time--reversed to the impinging electron.
If perfect coherence is maintained between electrons and
holes, an analogy may be drawn between the N--S
boundary and a phase--conjugating mirror, and one finds that
the hole retraces the possibly complicated motion of the
electron.  A weak magnetic field may break the symmetry
between electrons and holes, leading to large
effects in the conductance, of the order of $N \, e^2/h$.

Our purpose here is to suggest a new theoretical technique 
for calculating such effects in dirty mesoscopic 
N--S structures, which complements the 
existing theoretical tools in interesting ways.  
In this approach the system is described as a collection 
of ergodic scatterers, each possesing a random scattering
matrix.  The scattering cavities are connected to each 
other and to the external electrodes through leads of various 
widths.  In this first paper, we demonstrate the new approach 
by studying a system described by only one scattering cavity, 
attached to a single normal electrode and a single 
superconducting electrode through two ideal leads (see Fig.~1).  
The discussion of more complicated structures (having more 
than one scattering cavity, more than two electrodes, and 
non--ideal leads) is deferred to a future publication \cite{AZ}.

At first sight, it might seem that the conductances of such
systems would always be larger in the case with
electron--hole symmetry than in the non--symmetric case,
because whenever an electron happens to undergo Andreev
reflection, the charge transport is ``automatically''
doubled by the corresponding motion of the hole.  However,
only some of the partial waves representing an electron
entering the system from a normal electrode will end up
impinging on a superconducting electrode, and thus the hole
only imperfectly reproduces the wavefronts and the motion
of the electron.  Furthermore, the presence of Andreev
scattering actually reduces the density of states at the
Fermi surface --- an aspect of the proximity effect which
allows the gap of the superconductor to be manifest in the
adjacent normal material --- and thus the conductance may
sometimes be lowered more than it is enhanced.  In the
approach to be described below, the magnitude of both of 
these effects is described by a single quantity
$f$, which in principle may depend on the position within
the (normal part of the) system, and on the energy of the
pertinent electrons.  We define $f$ as the (averaged,
leading--order of the) probability amplitude for an
electron at that position and energy to propagate to a
superconducting electrode, be Andreev--reflected, and then
to coherently propagate back to the same position as a hole.
The density of states at the Fermi level is then equal to 
$(1-|f|^2)/(1+|f|^2)$ times the normal one \cite{AltZirn} 
(i.e.\ the density of states in the absence of induced 
superconductivity).

The present theoretical approach creates a link between the 
available calculational tools for disordered mesoscopic N--S 
structures.  On the one hand, the starting point
is the expression of the conductance in terms of scattering
properties (an appropriate Landauer formula), which
themselves are assumed to be described well by certain
ensembles of scattering matrices.  In this respect, the
approach is similar in spirit to a large number of recent
articles (e.g.\ those studying the 
Dorokhov--Mello--Pereyra--Kumar equation 
\cite{Been_RlT,Been_GBP,Brou,DMPK}), in
which the distribution of transmission--matrix eigenvalues
for various structures was studied and used to obtain the
conductance \cite{Frahm} (the transmission matrix is simply 
related to the scattering matrix of a structure, and carries 
the same information).  On the other hand, the large--$N$
diagrams used naturally lead to the definition of the
probability--amplitude $f$, which is closely related to the
space-- and frequency--dependent superconducting order
parameter of the Usadel equations \cite{Usadel}.  This set of
equations, which is widely used in the theory of dirty
superconductors, is ultimately based on the
Keldysh technique of diagrammatic perturbation theory.

Another approach which proves useful for disordered
mesoscopic systems is the semiclassical approach of the
Gutzwiller formula or the van Vleck formula \cite{sca}.
However, the semiclassical approximation for disordered
systems is not strictly unitary, and this turns out to be a
fatal flaw in the present context (see the next section).
Additional approaches developed for mesoscopic systems in
general, such as the nonlinear $\sigma$--model \cite{NLsM},
do not suffer from this flaw and are, in principle, applicable.

The outline of the paper is as follows.
The more technical part of the introduction is given in 
the following subsection.
In section II, we develop the formalism of planar
diagrams for averaging over large--$N$ random unitary matrices: 
the diagrammatic elements are discussed, and used
to form ``self--energies'' and ``effective couplings''.
This section is general, in the sense that its results are
used for both the simple junction studied in section III 
and the more complicated structures (networks of cavities) 
to be discussed in Ref.~\cite{AZ}.
In section III we proceed to apply the formalism to the 
simplest possible system --- that having only two leads
--- and compare our results with the literature.
Finally, conclusions and an outlook are given in section IV.

\subsection*{Preliminaries}

As discussed above, only normal--superconducting systems with 
simple geometries --- those describable by a single ``grain'' 
(Fig.~1) --- are considered in the present work.  The 
grain is taken to be an ideal scattering cavity with ergodic
single--electron internal dynamics, i.e., its scattering
matrix $S$ is assumed to be a random member of the
Circular Orthogonal Ensemble (COE) of random matrix theory
\cite{Mehta} (the ensemble of unitary symmetric matrices ---
the symmetry or ``orthogonality'' corresponds physically to
time--reversal symmetry).  This ensemble is known to
describe a wide universality class of physical systems,
including at least two different relevant categories:
clean cavities with shapes which lead to chaotic classical
dynamics, and disordered materials for which the chaotic
scattering is provided by impurities.  In both of these
cases one assumes that the escape time for an electron in
the cavity is much longer than the time required for
ergodicity --- otherwise the description using a
single random scattering matrix becomes unjustifiable.

The grain is connected to external electrodes via
leads which are assumed to be ideal (``waveguides'' of
electrons), and can have different widths, denoted $W_n$.
The physical widths are assumed to be much larger than
the Fermi wavelength, and the $W_n$'s are thus large integers 
which give the number of propagating modes in each lead.
The total number of modes, $N=\sum_n W_n$, gives the size
of the scattering matrix $S$.  In the simplest case, to be
discussed in detail in section III, there are only
two leads --- one ``superconducting lead'' of width $W_1$
connects the system to a superconducting electrode, and one
``normal lead'' of width $W_2$ connects it to a normal
electrode.  In this case, the $S$ matrix of the cavity is
written as
\begin{equation}
S=\pmatrix{ r & t' \cr t & r' \cr}  \; ,
\label{eq:S}
\end{equation}
where the reflection matrices $r$ (of size
$W_1 \times W_1$) and $r'$ (of size $W_2 \times W_2$) are
not equal in size, and the two transmission matrices $t$
and $t'$ are rectangular (except for the case $W_1=W_2$).  
In general, $S$ may have more than four blocks, describing 
scattering from and into more than two leads (at least one
normal lead and at least one superconducting lead are 
necessary for the conductances calculated here to be of 
relevance).  However, the statistical properties of the
$S$--matrix elements, according to the COE, do not
depend on the number and sizes of the blocks.

Once an electron has scattered in the cavity it may enter
the superconducting lead and impinge upon the
superconducting electrode.  As we will assume that the
electron is propagating at the Fermi level, it can not enter
the superconductor because of the gap, and it will be
completely reflected.  However, it may in principle undergo
normal reflection, and return as an electron, or Andreev
reflection \cite{Andreev} and return as a hole.  In the
present work we will assume for simplicity that the
probability of normal reflection vanishes, so that the
probability amplitude for Andreev reflection is of magnitude
unity (this assumption may be relaxed by introducing tunnel 
barriers in the ideal leads).  The phase of this probability
amplitude is determined from the Bogoliubov -- de Gennes
equations \cite{Slevin}, and the amplitude is found to be
$ie^{-i\phi}$ where $\phi$ is the phase of the
superconducting order parameter in the electrode.  The
scattering of holes in the cavity is described by $S^*$, the
complex conjugate of the scattering matrix for electrons.
The amplitude for a hole impinging on the N--S boundary to be
Andreev--reflected back into an electron is $ie^{i\phi}$.

The extra phase factors of $i$ are crucial to the description
of the induced proximity effects --- they lead to a relative
minus sign, i.e.\ destructive interference, between the
amplitude for an electron to reach any final state through a
certain history, and the amplitude to reach the same final
state through a history with two additional Andreev
reflections (into a hole and back into an electron).
In the absence of any time--reversal symmetry--breaking
effects, the dynamics is just such that two consecutive
Andreev reflections tend to reproduce the same state,
and so this destructive interference leads to a suppression
of (again) the {\it leading order} expression for the density of 
states, by the factor of $(1-|f|^2)/(1+|f|^2)$ mentioned 
above.

According to the appropriate Landauer formula \cite{Landauer},
the average conductance of such a structure (at zero
temperature and zero bias voltage) is given by
\begin{equation}
\left\langle  \sigma_{NS}  \right\rangle  \> = \>  {4e^2 \over h} \,
  \left\langle  {\rm tr} \> T_{eh} \, T_{eh}^{\dagger}  \right\rangle 
 \; ,
\label{eq:conductance}
\end{equation}
where $T_{eh}$ is an electron--hole ``transmission'' matrix,
giving the probability amplitudes for an electron injected
into the cavity through any one of the modes of the normal
lead to come back as a hole in any other of the modes of
that lead (the notation $\left\langle  \dots  \right\rangle$ 
denotes averaging of
the $S$ matrix over the ensemble).  This is similar to the
standard Landauer formula, 
$\sigma_N = (2e^2/h) {\rm tr} \, t \, t^\dagger$ 
where the factor of $e^2 / h$ is the quantum unit of 
conductance, i.e., the conductance per mode of an ideal lead 
with no scattering, the factor of $2$ comes from spin (which 
will play no role in this paper), and the trace sums the 
probabilities for electrons to be transmitted through the 
system.  The extra factor of $2$ in Eq.~(\ref{eq:conductance}) is 
due to the fact that for every electron from the normal lead 
that comes back as a hole, by charge conservation two 
electrons have gone into the superconductor.

The amplitudes in $T_{eh}$ are given by a sum over all
possible histories. In the simplest kind of history, an
electron injected into the cavity traverses the cavity with
the transmission amplitude $t'$, Andreev reflects off the
superconductor with the amplitude $i$ and becomes a hole,
and then the hole traverses the cavity with transmission
amplitude $t^{*}$. This gives a contribution of $t^{*}it'$.
As the system under discussion has only a single
superconducting lead, we have taken here for simplicity 
$\phi=0$, and avoided the extra phase factors (because of gauge 
invariance, they are only important in a system with at least 
two superconducting leads).  In the next simplest history, 
after Andreev reflection, the hole is scattered back towards 
the superconductor (rather than traversing the cavity), with
the reflection amplitude $r^{*}$. It then Andreev reflects
and becomes an electron. The electron bounces off the cavity
back towards the superconductor again with the reflection
amplitude $r$, only to come back as a hole after Andreev
reflection.  The hole finally traverses the cavity with
transmission amplitude $t^{*}$. This contributes an
amplitude of $t^{*}irir^{*}it'$ to $T_{eh}$.
The matrix $T_{eh}$ may thus be written as
\begin{equation}
T_{eh}=t^{*}(i+irir^{*}i+irir^{*}irir^{*}i+\dots)t'
\label{eq:T}
\end{equation}
This series for $T_{eh}$ could be written directly in terms
of the $S$ matrix --- for example, the term $t^{*}irir^{*}it'$
can be written as $i^3 S^{*}SS^{*}S$ --- provided that we
adopt the peculiar convention that in this product of four
unitary matrices the internal indices are to be summed
over only the $W_1$ indices pertaining to the lead
connecting the cavity to the superconductor and the
external indices are restricted to the $W_2$ indices
pertaining to the external lead.

Several simplifying assumptions are made here, such as
the complete absence of decoherence and symmetry
breaking effects, i.e.\ essentially zero applied voltage,
zero magnetic field, and zero temperature.  The results
are contrasted with those pertaining to the same system
with the symmetries completely broken, but the specific
form of the temperature or magnetic field dependence
cannot be evaluated so simply (as in Ref.\ \cite{Brou}), except
for the case of a magnetic field in a multiply--connected
geometry, with only simple Aharonov--Bohm phases.
Also, only the leading--order, $O(N)$, contribution to
the conductance is evaluated here, in contrast with the
transmission--matrix approach which often focuses
attention on the $O(N^0)$ corrections to the conductance
and its fluctuations \cite{Brou,DMPK}.  Another
drastic simplification is the description of possibly 
complicated experimental geometries by a single scattering
cavity, with the external leads considered as ideal.  As 
already mentioned, work on extending the present approach to 
deal with more realistic situations is in progress \cite{AZ}.

\section{
Planar diagrams for large $N$ unitary random matrices}

In this section we develop a diagrammatic technique which
may be used to evaluate the average conductance of
Eq.~(\ref{eq:conductance}), and is accurate to leading order in
$1/N$.  The diagrammatic description of the unaveraged
conductance is given in Fig.~2.  Propagation of an
electron through an ideal lead is denoted by a thin line
with an arrow to the right; propagation of a hole is
denoted by a line with a left--pointing arrow.  An
Andreev scattering event is denoted by a full semicircle.
Each element of $S$ carries two indices representing 
incoming and outgoing transverse modes in the leads, and
corresponds to a dangling double line.  These double lines
will eventually be connected to each other in various
ways, due to averaging over the ensemble of $S$.
The fact that one of the indices carried by two 
consecutive $S$--matrix elements is identical --- the mode 
number is not changed by Andreev reflection --- corresponds 
to the continuous single line which connects these two 
elements.  The conductance itself, according to 
Eq.~(\ref{eq:conductance}) and Eq.~(\ref{eq:T}), corresponds to 
a sum over all possible ``bubbles'' with two electron lines leaving 
the left--most point and two hole lines arriving at the
right--most point (see the figure), with an odd number of
Andreev reflections occurring along the lower line and along
the upper line.  We use the convention that the upper
line corresponds to the second $T_{eh}$ factor in
Eq.~(\ref{eq:conductance}), and all expressions pertaining to it
are thus complex--conjugated (or Hermitean--conjugated).

In the first subsection below, we give a general discussion
of averages of products of $S$--matrix elements, including
our use of time--reversal symmetry to simplify the
expressions, and the failure of a naive semiclassical
approach to the evaluation of these averages.
In the second subsection, we demonstrate that the
corresponding averages may be described diagrammatically
by introducing elements which couple $k$ copies of $S$ with
$k$ copies of $S^\dagger$, with weights as given in Fig.~3.
Although the correlations represented by the higher--order 
couplings of Fig.~3 are obviously weaker than those 
represented by the $k=1$, ``simple'' coupling, they affect 
more matrix elements, and thus contribute to the 
conductance of Eq.~(\ref{eq:conductance}) in the leading order.
It is remarkable that even though large random unitary 
matrices possess such more subtle correlations (when compared 
to the Gaussian ensembles of Hermitean matrices), an analogue 
of Wick's theorem still holds: a diagram 
composed of several of the elements of Fig.~3 carries a 
weight which, to leading order, is the product of the 
weights of the elements.

There is a close analogy to be drawn between the present 
diagrams and 't Hooft's analysis of quantum
chromodynamics in the large $N$ limit \cite{tHooft}:
the double lines are analogous to gluon propagators
and the single lines to those of quarks.  This analogy is 
useful because the index structure is similar --- each 
continuous single line carries an index which is to be 
summed over $O(N)$ different modes (and each double line 
carries the corresponding pair of indices).  Indeed, the 
basic rule that only ``planar'' diagrams contribute to the
leading--order results carries over to the present 
situation: a coupling with a weight of order $1/N^{2k-1}$ 
contributes to the leading order only if it divides the 
plane of the diagram into $2k$ disconnected sub--regions, 
thus increasing the number of factors of $N$ due to 
independent summations by $2k-1$.  Note that the analogy 
with quantum chromodynamics is certainly incomplete --- for 
example, the directions of the arrows in the diagrams is 
different, and there is no analogue here of gluon 
self--interactions (the couplings here connect $S$--matrix 
elements, and have no dynamics of their own).  
Another important difference is the fact that
the mode index here can belong to different leads, with
Andreev reflections occurring only in the superconducting
leads, and the external points (the left--most and
right--most points) imposing restrictions to modes in
normal leads --- each index is thus summed only over a subset 
of the $N$ possible values.

As often done in diagrammatic quantum field theory, we will
do the averaging in two steps.  First, we restrict ourselves
to connecting the double lines ($S$--matrix elements) within 
$T_{eh}$ to each other (and similarly the double lines 
within $T_{eh}^\dagger$ to each other),
i.e., we will sum over diagrams in which the upper and the
lower lines in the conductance bubble do not interact with
each other.  Later, we will connect the double lines within
$T_{eh}$ to those within $T_{eh}^\dagger$, and form
diagrams analogous to the diffuson contribution in the
calculation of the conductance of disordered metals
\cite{wellknown}.
The generic features of these two steps, involving the
``self--energy'' and the ``effective coupling'', will be
discussed in the third subsection below.  Specific
details which differ from one system to the
next are deferred to the following section.

\subsection{ Generalities }

One of the simplest examples of random--matrix ensemble 
averages to be considered is that of a product of only two 
$S$--matrix elements:
\begin{equation} \label{SSdag}
\left\langle  S_{ij} S^\dagger_{ji}  \right\rangle_{COE}  \> = \>
   {1 + \delta_{ij} \over N+1}  \; .
\end{equation}
This is an exact result --- valid for any $N$ (the
subscript COE implies that the system is time--reversal
invariant).  Summation over repeated indices is {\it not} 
implied in the present notation.  In the following, it will 
be important to distinguish between two ``leading--order''
approximations of this result, the ``naive'' one being
\begin{equation}
{1 + \delta_{ij} \over N+1}  \> \simeq \>
{1 + \delta_{ij} \over N}  \; ,
\end{equation}
and the other being
\begin{equation}
{1 + \delta_{ij} \over N+1}  \> {\stackrel{\sim}{=}} \>  
{1 \over N}
\end{equation}
(note the use of different relational operators).
Because each of the indices $i$ and $j$ is eventually to be 
summed independently over a set of a size of order $N$, the
$\delta_{ij}$ term will turn out to contribute less by
one factor of $N$, and is omitted in the ``true''
leading order.  In the present work, the notion of
``leading order'' is always used in this second sense,
unless the word ``naive'' is explicitly added.  In this 
example, a term which is naively of leading order disappears
in the eventual evaluation of the true leading order of
the conductance.  In contrast, below we will encounter terms
which are omitted in the naive leading order, but eventually
contribute to the leading order results because they 
contain less $\delta$--function factors.

For expressions involving arbitrarily many
$S$--matrix elements, the naive leading--order
results are easily found; for example they can be obtained
from a semiclassical approximation \cite{sca}.
In this approximation $S$ is written as
\begin{equation} \label{scaS}
S_{ij}  \> \simeq \>
\sum_{\mu \in \{ ij \} }  A_\mu e^{i s_\mu/\hbar}  \; ,
\end{equation}
where $\mu$ is an index enumerating classical orbits in
the set $\{ ij \}$, i.e.\ those that have initial conditions
corresponding to mode $j$ and final conditions
corresponding to mode $i$ (classical orbits which enter 
and leave the cavity at angles specified by
$j$ and $i$ respectively, and propagate at the Fermi energy).
The amplitude for propagation along the orbit $\mu$ is
denoted by $A_\mu$, and its classical action by $s_\mu$.

Products of elements of $S$ correspond to multiple sums
over orbits, and averaging corresponds to dropping all terms
which have non--trivial phase factors \cite{Berry}.  The
approximation here is simply that $\hbar$ is small relative
to the possible fluctuations in the actions $s_\mu$,
regardless of whether these fluctuations occur within a certain
set $\{ ij \}$ due to the classically chaotic nature of the
dynamics in the cavity, or between different realizations
of a disordered potential in the cavity.  Consider, for example, 
the case without time--reversal symmetry, denoted by the 
subscript CUE (Circular Unitary Ensemble).  In this case we 
have
\begin{equation} \label{scaSSd}
\left\langle  S_{ij} S^\dagger_{ji}  \right\rangle_{CUE} \> \simeq \>
    \left\langle   \sum_{\mu,\nu \in \{ ij \} }
           A_\mu A_\nu^* e^{i(s_\mu-s_\nu)/\hbar}  \right\rangle
 \> \simeq \>
  \left\langle  \sum_{\mu \in \{ ij \} } |A_\mu|^2 \right\rangle
 \> = \> {1 \over N}  \; .
\end{equation}
The last equality follows from the correspondence between
$|A_\mu|^2$ and classical probabilities, and the assumption
of ergodicity, i.e.\ that the particle ``forgets'' its
initial conditions once it enters the cavity, and is
equally likely to leave the cavity in any outgoing mode
or direction.

For products of many elements of $S$ and $S^\dagger$, the
obvious generalization is to pair each element of $S$ with
an element of $S^\dagger$ in all possible manners, so that
cancelations of the actions can occur.  For example,
\begin{equation} \label{SSSS}
\left\langle  S_{ij} S^\dagger_{jk} S_{kl} S^\dagger_{li}  
\right\rangle_{CUE}  \> \simeq \>  {1 \over N^2}
 \left( \delta_{ik} + \delta_{jl} 
\vphantom{\mbox{$1\over 2$}} \right) \; .
%\LP \delta_{il} \delta_{jk} \delta_{mp} \delta_{no} +
% \delta_{ip} \delta_{jo} \delta_{ml} \delta_{nk} \RP  \; .
\end{equation}
Such expressions may be obtained by joining the double
lines in the diagrams using only the first element of 
Fig.~3, as displayed in the first few diagrams in Fig.~4 
[specifically, that of Fig.~4(a), the first two in 
Fig.~4(b), and the first five in Fig.~4(c); in the last 
case the sixth possible permutation is omitted because it 
contributes at a higher order in $1/N$].
This corresponds to ignoring the unitarity of the matrices,
and taking the results for Hermitian matrices instead (the
Gaussian Unitary Ensemble). 
As we will see in the next subsection, an additional term of
order $1/N^3$ must be added to the right hand side of this
equation in order not to violate unitarity, and such 
``naively'' higher--order terms eventually affect the leading 
order of the conductance.

In the case with time--reversal symmetry (COE), the
action differences $s_\mu-s_\nu$ vanish identically not
only for $\nu$ equal to $\mu \in \{ ij \}$, but also for
$\nu$ equal to $\mu^T \in \{ ji \}$, which is the orbit
related to $\mu$ by time reversal.  This leads to an extra
factor of $1+\delta_{ij}$ in the second--to--last and last
expressions in Eq.~(\ref{scaSSd}), and similarly to additional
terms in Eq.~(\ref{SSSS}).  However, the ``true'' leading order
result is independent of the symmetry, e.g.,
\begin{equation} \label{coe_cue}
\left\langle  S_{ij} S^\dagger_{ji}  \right\rangle_{COE}  
\> \stackrel{\sim}{=} \>
\left\langle  S_{ij} S^\dagger_{ji}  \right\rangle_{CUE}  \; .
\end{equation}
The same holds also for averages involving arbitrarily
many matrix elements alternating from $S$ and $S^\dagger$.
For the naive leading order contribution to the ``true''
leading order, this follows directly from the semiclassical
approximation discussed above, as can be seen by imagining
all possible ways of connecting all the matrix elements
with simple double lines --- whenever we use the symmetry of
$S$ to generate additional terms, we break the planarity of
the diagram (the corresponding double line is crossed) and 
thus generate a contribution of a ``true'' higher order 
(containing additional $\delta$--function factors).  In 
the next subsection we will see that the
terms which are of both naive and ``true'' leading
order, together with the property of unitarity, suffice to
determine the terms which are naively of higher order but
contribute to the leading order of the conductance.  Thus,
as both the COE and the CUE are ensembles of unitary
matrices, Eq.~(\ref{coe_cue}) and its many--$S$--and--$S^\dagger$
generalization must hold also beyond the naive leading order.
Note that it is of course crucial to use $S^\dagger$ and not 
$S^*$ in Eq.~(\ref{coe_cue}).  For example
$\left\langle  S_{ij} S^*_{ji}  \right\rangle_{CUE} = \delta_{ij}/N$ 
and would contribute in a higher order than
$\left\langle  S_{ij} S^*_{ji}  \right\rangle_{COE}$ 
given in Eq.~(\ref{SSdag}).

Below we will use the time--reversal symmetry property,
$S^T=S$ and $S^*=S^\dagger$, to rewrite all our expressions
in terms of $S$ and $S^\dagger$ only (they naturally appear
in alternating order), and then rely on Eq.~(\ref{coe_cue}) to
avoid the necessity of actually computing in the COE.
For example, take the term 
$\left\langle S^*_{kj} S_{ji} \left( S^*_{kn} S_{nm} S^*_{ml} S_{li} 
\vphantom{\mbox{$1\over 2$}} \right)^*  \right\rangle_{COE}$ 
displayed in Fig.~2, which describes interference between the 
amplitude for a hole produced by a single Andreev 
reflection and that produced by three consecutive Andreev 
reflections (with the multiplicative factors of $i$ 
omitted).  This term is first transformed to read
$\left\langle  S^\dagger_{kj} S_{ji} S^\dagger_{il} S_{lm} 
S^\dagger_{mn} S_{nk} \right\rangle_{COE}$,
and then the leading order in $N$ result for its average
is obtained within the CUE.
Such averages over the unitary ensemble, i.e.\ over the
unitary group $U(N)$, have been performed, e.g., in
Ref.~\cite{RMTave}, for general $N$.  We proceed to follow
a somewhat heuristic diagrammatic approach for evaluating
them to leading order in $1/N$, referring to this work for 
a mathematically rigorous derivation.

If time--reversal symmetry is broken, the replacement of 
$S^*$ by $S^\dagger$ is not permissible, and this affects
the leading order results for the conductance, as
already emphasized in the introduction.
The symmetry may be broken by giving the electrons and
holes a finite excitation energy, due to either a finite
bias voltage or a finite temperature.  In this case,
the electrons travel with an energy $E_F + \epsilon$ and
the corresponding holes with an energy of $E_F -\epsilon$
(where $E_F$ denotes the Fermi energy and $\epsilon$ the
positive excitation energy).  Thus $S^*$ is in fact the
complex conjugate of $S$ evaluated at a {\it different}
energy, and the conductance may be obtained by assuming
that $S$ and $S^*$ are two {\it independent} members of
the CUE, if $\epsilon$ is greater than the range in
energy over which correlations in $S$ die out.
This range or correlation energy, also called the Thouless
energy, is easily evaluated from Eq.~(\ref{scaS}): one notes that
the derivative of the classical action $s_\mu$ with
respect to the energy is equal to $t_\mu$, the duration of
the orbit $\mu$.  All of these orbits are of lengths of
the order of $t_{esc}$, the escape time from the cavity,
and thus the Thouless energy is equal to $\hbar/t_{esc}$.

Another effect which breaks the symmetry between electrons
and holes involves applying a magnetic field, stronger
than the corresponding correlation field \cite{corB}.
Here $S^*$ is the complex conjugate of $S$ (evaluated at
the same energy), but is distinct from $S^\dagger$.
However, the leading--order results for the
conductance may still be obtained by considering
them as {\it independent} members of the CUE (see,
e.g., Ref.~\cite{Brou}).  This can easily be demonstrated
using the diagrammatic language ---
all the planar diagrams have in this case the property
that all the double lines connect $S$'s to $S^\dagger$'s and
$S^*$'s to $S^T$'s, with no intermixing [the argument is
essentially the same as that used in the discussion of
Eq.~(\ref{coe_cue})].
In the physical discussion of the following section,
the results for the symmetric case are compared with
those pertaining to such time--reversal non--symmetric
situations, which turn out to be much easier to evaluate 
(cf.\ the last paragraph of this section).

\subsection{Generalized couplings for unitary matrices}

We thus would like to calculate expressions of the type
$\left\langle S_{ia}S^{\dagger}_{aj}S_{jb}S^{\dagger}_{bk} \dots
S_{lc}S^{\dagger}_{ci}  \right\rangle$, where 
$\left\langle \dots \right\rangle$ denotes the
average over the unitary group $U(N)$.  We have used letters
from different sections of the alphabet to emphasize the fact
that, for elements of this group, the first (or upper) index
and the second (or lower) index of $S$ transform differently
(covariantly and contravariantly) under $U(N)$
(the reverse holds for $S^{\dagger}$).
It follows directly from the invariance of the probability
distribution of $S$ under such $U(N)$ transformations
that the averaged product of any $n$ elements of $S$ and $m$
elements of $S^\dagger$ may be written as 
\begin{eqnarray}
&& \left\langle  S_{i_1 a_1} S_{i_2 a_2} \dots S_{i_n a_n} \>
S^\dagger_{b_1,j_1} S^\dagger_{b_2 j_2} \dots
S^\dagger_{b_m j_m} \right\rangle  \> = \>  
\nonumber \\ && \qquad\qquad\qquad
\delta_{nm} \sum_{u,v}
C_{u,v} \> \delta_{i_1 j_{u(1)}} \delta_{i_2 j_{u(2)}}
\dots \delta_{i_n j_{u(n)}} \>  \delta_{a_1 b_{v(1)}}
\delta_{a_2 b_{v(2)}} \dots \delta_{a_n b_{v(n)}}  \; ,
\end{eqnarray}
where $u$ and $v$ are permutations of $n$ elements and
$C_{u,v}$ are coefficients to be determined (the
average vanishes of course for $m$ different from $n$).
Thus, $\delta$--functions such as $\delta_{ia}$ or
products of them will never appear.

These products of $\delta$--functions may be represented
diagrammatically using the elements of Fig.~3, with a simple
$k \! = \! 1$ coupling for every 1--cycle in the combined 
permutation $u v^{-1}$, a four--sided ($k \! = \! 2$) coupling 
for every 2--cycle, etc.
The condition of planarity of the diagrams does not enter at
this stage because we have not yet distinguished between
contributions of different orders.
The purpose of this subsection is to demonstrate that the
leading--order expressions for the $C_{u,v}$ coefficients
can be expressed as products of weights corresponding to
the individual cycles in $u v^{-1}$, and that these weights
are indeed those given in Fig.~3.  Interestingly, we show
that all the necessary properties of the $C_{u,v}$'s follow
from the unitarity of $S$, i.e.\ from the equalities
$\sum_a S_{ia} S^\dagger_{aj} = \delta_{ij}$ and
$\sum_i S^\dagger_{ai} S_{ib} = \delta_{ab}$.
The unitarity of $S$ corresponds physically to the
conservation of the number of particles in the cavity,
i.e.\ to the fact that the incoming flux of electrons (or
of holes) is equal to the outgoing flux.

We will now calculate the required averages iteratively,
starting from averages of only two elements and gradually
building up.  Start with
\begin{equation}
\left\langle  S_{ia}S^{\dagger}_{ai} \right\rangle  \> = \>
{1 \over N} \equiv {\gamma_2 \over N}  \; .
\end{equation}
The equality here follows from the unitarity of $S$ as we can
see by summing both sides of the equality over $a$ (or $i$).
This is represented in figure 4(a) (note that in this figure 
the arrows have been omitted, and the matrix elements appear 
in counter--clockwise, rather than clockwise, order).  
We have introduced a quantity $\gamma_2$ defined to be $1$ for 
later convenience.

We next write
\begin{equation}
\left\langle  S_{ia}S^{\dagger}_{aj}S_{jb}S^{\dagger}_{bi} \right\rangle 
\> \stackrel{\sim}{=} \>
{ \delta_{ij}+\delta_{ab} \over N^2} + {\gamma_4 \over N^3}
\label{eq:gamma4}
\end{equation}
where the first term corresponds to the naive leading order
result, and the second term is a possible correction
of naively higher order.  It involves $\gamma_4$, which is an
unknown ``coupling constant'' of a four--sided interaction 
[see Fig.~4(b)]. To determine $\gamma_4$ we sum
this equation over $b$.  The left hand side gives
$\delta_{ij} \left\langle  S_{ia}S^{\dagger}_{ai} \right\rangle 
= {\delta_{ij} \over N}$,
the right hand side
${(1+ \delta_{ij} N+\gamma_4)\over N^2}$,
and thus we obtain $\gamma_4 = -1 = -\gamma_2\gamma_2$.

To see the recursion pattern, we go one step further and
consider
\begin{equation}
\left\langle  S_{ia}S^{\dagger}_{aj}S_{jb}S^{\dagger}_{bk}S_{kc}
S^{\dagger}_{ci} \right\rangle  \> \stackrel{\sim}{=} \>  (\dots\dots) +
{\gamma_2 \gamma_4 \over N^4} \left( \delta_{ac}+\delta_{ik}+
\delta_{bc}+\delta_{jk}+\delta_{ab}+\delta_{ij} 
\vphantom{\mbox{$1\over 2$}} \right) +
{\gamma_6 \over N^5}  \; ,
\label{eq:gamma6}
\end{equation}
see Fig.~4(c).  Here $(\dots\dots)$ represents the naive 
leading order term --- a sum of terms each involving two 
Kronecker deltas, for example,
$\delta_{ab}\delta_{bc}/N^3$ or $\delta_{ik}\delta_{ab}/N^3$
--- which are generated by using only the ``simple''
$\gamma_2$ couplings.
The second term, involving a single $\delta$ function, is due
to diagrams with one ``simple'' coupling, $\gamma_2$, and one
four--sided vertex, $\gamma_4$.  The third term corresponds to
the six--sided vertex which must be introduced as well.

One can easily see that the terms required by unitarity at (say)
the $c$ vertex are correctly reproduced by all the diagrams with
a simple double line connecting $S_{kc}$ with
$S^\dagger_{ci}$, and all the previously derived elements (in
this case $\gamma_2$ and $\gamma_4$) connecting the other four
$S$--matrix elements.  Of the twelve terms in Fig.~4(c), only
three are of this type, and the other nine terms must thus
cancel each other when summed over $c$.
There are two types of cancelations between these ``unwanted''
terms.  The first type occurs, for example, between the term
$\delta_{ab}\delta_{bc}/N^3$ and the term $-\delta_{ab}/N^4$.
More generally, such cancelations occur due to the weights of
the lower--order couplings already determined (in
this case the weight $\gamma_4$), with a ``distant'' part
of the diagram (in this case the simple coupling leading to
the $\delta_{ab}$ factor) playing only a passive role.  There
are three pairs of diagrams in Fig.~4(c) which cancel in this
way.  The second type of cancelation accounts for the
remaining three diagrams, including the one with the new
six--sided vertex, whose weight $\gamma_6$ we need to
determine.  The diagram with the six-sided vertex involves 
no $\delta$--functions, and the only terms which upon
summing over $c$ would produce no Kronecker delta are
evidently
the two terms $\delta_{bc}+\delta_{ca}$. Thus, we have the
recursion relation $\gamma_6=-2\gamma_2 \gamma_4$.

Speaking picturesquely, we see that we have a
combinatorial problem in which in general $2k$ diners are
seated around a round table and engaging in conversation.
Note that the indices $i, a, j, .....$ are to be associated
with the spaces between the diners.  Each diner is identified
by two indices: thus, diner $[ia]$, whom we may think of as a
gentleman, is associated with $S_{ia}$, diner $[aj]$, whom we
may think of as a lady, is associated with $S^{\dagger}_{aj}$,
and so forth. The number of diners $2k$ is required to be 
always even.  In the case $2k=6$ we just discussed,
$\gamma_6$ corresponds to all six diners engaging in one
conversational group, $\gamma_4 \gamma_2$ corresponds to four
of the diners engaging in one conversational group with the
other two speaking only to each other. The rules of large $N$
tells us that etiquette dictates that the ``lines of
conversation'' cannot cross.

Proceeding in this way, we see that for the general case the
second type of cancelation will occur if
\begin{equation}
\gamma_{2k} + \sum_{l=1}^{k-1} \gamma_{2l} \gamma_{2(k-l)}
\> = \>  0  \; .
\label{eq:recursion}
\end{equation}
We can readily interpret the terms as follows: in the first
term all of the $2k$ diners are engaged in one conversation,
while in the second term the group has broken up in two,
with the two diners who share the index on which we are
summing (which was index $c$ in the above example) belonging
to different groups.  The sum is over all terms with a single
$\delta$ function connecting that index ($c$) with all others
of its type (covariant or contravariant).  There are always
an even number of diners in any conversational group, the
number of ``males'' being equal to the number of ``females''
--- the subscript of $\gamma$ is always even.

This last fact clearly suggests defining
$c_j=(-1)^{j+1}\gamma_{2j}$,
in terms of which we have the recursion relation
\begin{equation}
c_k  \> = \>  \sum_{j=1}^{k-1} c_j c_{k-j}  \; ,
\label{eq:rec}
\end{equation}
with the explicit  solution
\begin{equation}
c_k  \> = \>  {(2k)! \over 2(2k-1)(k!)^2} 
\label{eq:cks}
\end{equation}
[another way of stating this result is: $c_k = (4-6/k) c_{k-1}$; 
$c_1=1$].
Actually, for our later applications we don't need the explicit
solution for $c_k$ as much as the generating function defined
by $w(x)=\sum_{k=1}^{\infty} c_k x^k$.  The recursion relation
above immediately implies
\begin{equation}
w  \> = \>  x+w^2  \; ;
\label{eq:gen}
\end{equation}
the extra term $x$ on the right hand side can be seen
by noting that the series for $w(x)$ starts with $x$ for small
$x$ and hence $w^2$ starts with $x^2$.  We thus directly
obtain
\begin{equation}
w  \> = \>  \left( 1- \sqrt{1-4x} 
\vphantom{\mbox{$1\over 2$}} \right) /2  \; ,
\label{eq:w}
\end{equation}
which implies (\ref{eq:cks}).  In the appendix, we give an
alternative derivation, which involves generalizing the
energy--dependent Green's functions often used for random
Hermitean matrices to the case of unitary matrices.

In the above, we have tacitly assumed that the weights of
diagrams with several coupling elements are given by
the product of the weights of the elements (those of Fig.~3).
Our argument so far does not demonstrate that this is the
only way to maintain unitarity.  However, the unitarity
relations between diagrams with $2k$ matrix elements and with
$2k-2$ matrix elements do in fact suffice to determine the
weights uniquely \cite{RMTave}, and therefore the weights we
have found are the correct weights.  To see this, imagine
that all weights of all diagrams with $2k-2$ or less matrix
elements have been determined, and that you need to
determine the weight of the diagrams with $2k$ elements.  
Start from all diagrams having at least one simple double 
line connecting two adjacent ``diners'' or matrix elements.  
By summing over the index shared by these two diners, one 
immediately gets the weight of that diagram as $1/N$ times 
the weight of the corresponding diagram with that double 
line omitted (as well as the two diners it connects).  
For diagrams which do not have
such pairs, one can start from a foursome of adjacent diners,
and sum over any one of the indices internal to that
foursome.  Obviously this can be continued until all diagrams
of $2k$ elements have been accounted for, with the last
diagram serving to determine $\gamma_{2k}$ according to
Eq.~(\ref{eq:recursion}).
Note that the actual $C_{u,v}$ for general $N$ have more
complicated expressions, with special cases for $n > N$.
However, in the large $N$ limit, corrections to the above
analysis (e.g.\ due to non--planar diagrams) are smaller
than the leading term by a factor of $1/N^2$.  This high
degree of accuracy of the leading order is specific to
the case without time--reversal symmetry --- corrections of
relative order $1/N$ would be present in the COE.

\subsection{The self--energy $\Sigma$ and the effective
coupling $\Gamma$}

According to the diagrammatic rules established above,
only planar diagrams contribute to the leading order
results.  In the present subsection we use these rules to
define a ``self--energy'' and an ``effective coupling''
in a manner analogous to similar quantities which appear
in diagrammatic analyses of perturbative field theory.
We thus group together diagrams which differ from each
other in their internal structure, but not in the way
they connect to the rest of the diagram (or the rest of
the ``plane'').  There are two relevant families --- the
``one--particle irreducible'' diagrams are summed in
Fig.~5(a), and the ``two--particle irreducible''
diagrams are summed in Fig.~5(d).  These elements
may then be used to calculate the ``averaged Green's
function'' $G$, the ``diffuson'' $D$, and eventually the
conductance $\sigma$ (see Fig.~5).
Note that although the analogy to the usual perturbation
theory of disordered metals is both striking and useful,
there are also important differences: the scattering is
included in a very different way, and unlike a typical
Green's function in quantum field theory, there is no
energy variable involved in $G$ or $\Sigma$ (at least at
the present level).

Following from Fig.~5, we have a ``Schwinger--Dyson'' 
equation for the Green's function
\begin{equation}
G  \> = \>  G_0 + G_0 \Sigma G  \; ,
\label{sd}
\end{equation}
or equivalently
\begin{equation}
G^{-1}  \> = \>  G_0^{-1} - \Sigma  
\label{eq:sd1}
\end{equation}
(from here on we use $=$ rather than 
$\stackrel{\sim}{=}$, although it should be remembered 
that results are being evaluated only to leading order).
The notation here is such that $G_0$ represents
propagation with a single Andreev reflection
\cite{notation} --- it is a two--by--two,
off--diagonal matrix in electron--hole space,
whereas it is diagonal in mode--space.
Similarly, $\Sigma$ and $G$ are off--diagonal in
electron--hole space.  Note that the self--energy $\Sigma$
differs from the other quantities in this equation
in that it is proportional to the unit matrix in
mode space --- the diagrams of Fig.~5(a) are
independent of the (single) mode index carried by
the external lines.

It is convenient to parameterize $\Sigma$ by a single complex
parameter $f$,
\begin{equation} \label{parame}
\Sigma = \pmatrix{ 0 & if^* \cr if & 0 \cr}  \; ,
\end{equation}
where the structure in mode space is not explicitly
written.  The fact that the electron--hole element $if$ is
minus the complex conjugate of the hole--electron element
$if^*$ follows directly from the fact that a single Andreev
reflection, and thus $G_0$ and $G$, also share this structure.
One may refer to $f$ as the amplitude for Andreev
reflection in the cavity, because following from Fig.~5(a),
to leading order in $N$ and for the averaged behavior,
an electron entering the cavity from any one of the leads
may be considered to be Andreev reflected directly into the
time--reversed hole trajectory, with a probability amplitude
$f$ (and the extra phase factor $i$).  

The Green's functions $G_0$ and $G$ have a more complicated 
internal structure.  To maintain the generality of the 
present section, we consider here a cavity connected to 
{\it several} superconducting and normal leads.  The bare 
Green's function, $G_0$, is parameterized as in Eq.~(\ref{parame}), 
with $f$ replaced by $e^{-i\phi}$, using different values of 
$\phi$ for modes in different superconducting leads, and $f$ 
replaced by 0 (and hence $G_0=0$) for modes in normal
leads.  The corresponding elements of $G$ are,
according to Eq.~(\ref{eq:sd1}), similarly described by a parameter
$g = 1/(e^{i\phi}+f^*)$, and since $\phi$ varies from one 
superconducting lead to the next, so does $g$.  This total 
amplitude for transforming an electron into a hole can be 
understood as a sum over terms with an odd number of Andreev 
reflections:
$ig = ie^{-i\phi} + i^3 e^{-i\phi} f^* e^{-i\phi} +
   i^5 e^{-i\phi} f^* e^{-i\phi} f^* e^{-i\phi} + \dots$.

The specific values of $g$, and the number of modes for
which each one of them occurs, depend on the geometrical
structure of the system.
Such specifics will be discussed in the following section.
However, one may make progress by introducing a notation
for the trace of each of the off--diagonal blocks of $G$,
which reads $i N \alpha = \sum_j  G_{j,h;j,e}$, and
leads directly to
\begin{equation} \label{eq:gamma}
\alpha = \sum_n {W_n \over N} g_n =
\sum_n {W_n \over N} {1 \over e^{i\phi_n}+f^*}  \; ,
\end{equation}
where the sum over $n$ runs over the superconducting
lead(s).  The trace of the hole--electron block is similarly
equal to $i N \alpha^*$.  We proceed to derive a general
relationship between $\alpha$ and $f$, which taken together
with Eq.~(\ref{eq:gamma}) will eventually allow us to find $f$ for
particular systems.

According to Fig.~5(a), the ``self energy'' $\Sigma$
is given by a sum of terms involving higher and
higher order couplings.  Each $\gamma_{2k}$ coupling 
divides the plane of the diagram into an external part and 
$2k-1$ internal ``triangles'', which give rise to factors 
of $\alpha$ and $\alpha^*$ (traces over $G_{he}$ and 
$G_{eh}$).  Mathematically, this relationship reads
\begin{equation}
f = \gamma_2 \alpha - \gamma_4 \alpha |\alpha|^2
+ \gamma_6 \alpha |\alpha|^4 + \dots
=\sum_{k=1}^{\infty} \gamma_{2k} \alpha \left( -|\alpha|^2 
\vphantom{\mbox{$1\over 2$}} \right)^{k-1}
\; ,
\label{eq:sigma}
\end{equation}
or, in terms of the $c_k$s,
\begin{equation}
f =  \sum_{k=1}^{\infty} c_k \alpha |\alpha|^{2k-2}  \; .
\label{eq:seriesforf}
\end{equation}
Comparing with (\ref{eq:w}) we find
\begin{equation}
f={1\over 2\alpha^*} \left( 1-{\sqrt {1-4|\alpha|^2}} 
\vphantom{\mbox{$1\over 2$}} \right)  \; .
\label{eq:f}
\end{equation}
Actually, a more useful form comes directly from
Eq.~(\ref{eq:gen}) with the substitution
$w \rightarrow \alpha^* f$ and
$x \rightarrow |\alpha|^2$ (using the fact that the 
complex phase of $f$ is equal to that of $\alpha$):
\begin{equation}
\alpha={f \over {1+|f|^2} }  \; ,
\label{eq:alpha}
\end{equation}
a result which we will use repeatedly.

We now turn to a similar consideration of the
``two--particle irreducible'' diagrams, i.e.\ the
``effective coupling'' $\Gamma$ depicted in Fig.~5(d), and its
relationship with the ``diffuson'' $D$ of Fig.~5(e).  The
structure in mode space is again trivial, with both $\Gamma$
and $D$ having two independent mode indices on the left and
right (in fact one should consider $\Gamma$ and $D$ to have
four mode indices, with the right two and the left
two being equal to each other).  However, the structure in
electron--hole space is more complicated than before,
having both diagonal and off--diagonal elements.  Before
any summation over the external mode indices, the diagrams
in Fig.~5(d) and Fig.~5(e) are of order $1/N$, and we define
$\Gamma$ and $D$ to be $N$ times the corresponding
diagrams.  We parameterize
$\Gamma$ by $\Gamma= (\Gamma_d I+ \Gamma_c \tau)$, where $I$
(often omitted) and
$\tau = {\mbox{\small $0 \; 1$} \choose \mbox{\small $1 \; 0$}}$
denote the identity matrix and the first Pauli matrix
respectively.  Here $\Gamma_d$ (for ``diagonal'') is the
probability for an electron to remain an electron, and a hole
to remain a hole, while $\Gamma_c$ (for ``change'') is the
probability for an electron to become a hole and vice versa.
Similarly, we write $D= (D_d I+ D_c \tau)$.
According to the diagrams of Fig.~5(e), we have
\begin{equation}
D  \> = \>  \Gamma + \Gamma \, t \, D
\label{eq:eqforD}
\end{equation}
or
\begin{equation}
D^{-1}  \> = \>  \Gamma^{-1} - t  \; ,
\label{eq:Dinverse}
\end{equation}
where
\begin{equation}
t  \> = \>  \sum_n {W_n\over N} \, |g_n|^2 \, \tau
\label{eq:t}
\end{equation}
represents the two Green's functions $G$ and $G^*$
connecting $\Gamma$ to $D$.  The $\sum_n W_n$ appears
explicitly in our two--by--two matrix notation, whereas it 
would be implicit in the matrix multiplication in 
Eq.~(\ref{eq:eqforD}) if multiplication in the mode sub--space were 
implied [the factor of $1/N$ is introduced in $t$ to account 
for our definition of $\Gamma$ and $D$ as $O(N^0)$ objects].
Again, the specific values for $t$ and $D$ depend very much
on the geometric structure, and are left for the following
section.  Here we proceed to derive the general
relationship between $\Gamma$ and $\alpha$ (and therefore 
$f$), which, as in the case of the self--energy $\Sigma$, 
is the only aspect of $G$ which enters in the definition of
$\Gamma$.

Let us then look at the graphs contributing to $\Gamma$,
Fig.~5(d), in particular the terms with $2k$ double lines 
joined by a coupling constant $\gamma_{2k}$.  Those graphs 
in which an odd number of double lines land on the top line 
(and thus necessarily with an odd number of double lines 
landing on the bottom line) contribute to $\Gamma_d$, and 
not to $\Gamma_c$.  There are $k$ such possibilities (one 
double line can land on the top line, three double lines, and 
so on, up to $2k-1$ double lines).  In contrast, those graphs 
in which an even number of double lines land on the top and 
bottom lines contribute to $\Gamma_c$.  There are $k-1$ such
possibilities.  We thus find that
\begin{equation}
\Gamma_d  \> = \>  \sum_{k=1}^{\infty} c_k |\alpha|^{2k-2} k
\; ,
\label{eq:gammadseries}
\end{equation}
where in each contribution to $\Gamma_d$ the signs in 
$\left( -|\alpha|^2 \vphantom{\mbox{$1\over 2$}} \right)^{k-1}$ 
cancel with the sign of 
$\gamma_{2k}$.  Comparing with (\ref{eq:seriesforf}) we see 
that $\Gamma_d = {d(\alpha^*f) \over d(|\alpha|^2)}$, and thus
\begin{equation}
\Gamma_d  \> = \>  {1 \over \sqrt{1-4|\alpha|^2} }
          \> = \>  { 1+|f|^2 \over 1-|f|^2}
\label{eq:gammad}
\end{equation}
from Eqs.~(\ref{eq:f}) and (\ref{eq:alpha}).  Similarly, we 
have
\begin{equation}
-\Gamma_c = \sum_{k=1}^{\infty} c_k |\alpha|^{2k-2} (k-1)
\; .
\label{eq:gammacseries}
\end{equation}

Note the minus sign in the last equation.  To see its origin, 
consider the term involving four double lines with two 
landing on the top single line and two landing on the bottom 
line: it is given by
$\gamma_4 (i\alpha)(-i\alpha^*)$ for the electron--hole term, 
and $\gamma_4 (i\alpha^*)(-i\alpha)$ for the hole--electron 
term (they are equal; recall that expressions pertaining to 
the top line are complex--conjugated).  Due to the 
negative sign of $\gamma_4$, this gives a negative 
contribution to $\Gamma_c$.  Next, consider a typical term 
involving six double lines, say with two double lines going 
upwards and four double lines going downwards. 
The contribution to $\Gamma_c$ is given by, say,
$\gamma_6 ( i^3 \alpha |\alpha|^2)(-i\alpha^*)$, but
$\gamma_6$ is positive. Proceeding in this way, we see
that $\Gamma_c$ is given by a series with negative
coefficients.  In contrast, $\Gamma_d$ is given by a
series with positive coefficients. Consider for example
the term just mentioned with six double lines. 
To obtain a contribution to $\Gamma_d$ we have
to move, for example, one double line from the bottom 
to the top part of the diagram. In the expression just given,
one factor of $i\alpha$ is thus changed to $-i\alpha$.
As already emphasized in the introduction, various signs
of this type, and the factors of $i$ which generate them,
are of crucial importance to obtaining our final results,
and reflect fundamental aspects of the physics involved.

To continue with our evaluation of $\Gamma_c$ we obtain
by its defining series (\ref{eq:gammacseries})
that
\begin{equation}
-\Gamma_c  \> = \>  \Gamma_d - {f \over \alpha}
 \> = \>  { |f|^2 (1+|f|^2) \over 1-|f|^2 }  \; .
\label{eq:gammac}
\end{equation}
Putting $\Gamma_d$ and $\Gamma_c$ together and
evaluating $\Gamma^{-1}$, we find the remarkably simple
formula
\begin{equation}
\Gamma^{-1}  \> = \>
 {\Gamma_d-\Gamma_c \tau \over \Gamma_d^2 -
\Gamma_c^2}
 \> = \>  {1\over (1+|f|^2)^2} (1+|f|^2 \tau)  \; .
\label{eq:gammainverse}
\end{equation}
As we saw from (\ref{eq:Dinverse}), we need the inverse of
$\Gamma$, rather than $\Gamma$ itself.

In the following section, we simply apply these
formulas --- Eqs.~(\ref{eq:gamma}), (\ref{eq:alpha}),
(\ref{eq:Dinverse}), (\ref{eq:t}) and
(\ref{eq:gammainverse}) --- to a specific geometry, 
and proceed to evaluate diagrammatically the average 
conductance of Eq.~(\ref{eq:conductance}).

Before continuing, we mention how 
the procedure is changed when one wishes to evaluate the 
conductance in the absence of time--reversal symmetry.  
According to the discussion at the end of subsection A 
above, in this case the leading order contributions 
cannot contain couplings connecting electron scattering 
elements and hole scattering elements along the bottom 
line (or the top line) of the diagrams.  
Thus there are no contributions to $\Sigma$ 
(to leading order), and we find that $f=0$ 
in this case (and thus also $\alpha=0$).  Similarly, only 
the first ``simple'' diagram in Fig.~5(d) is a 
legitimate contribution, and $\Gamma = \Gamma^{-1} = I$.  
Lastly, we have $G=G_0$ in this case, and 
$|g_n|=|e^{-i\phi_n}|=1$ should be used in Eq.~(\ref{eq:t}).
In order to obtain the conductance, one still needs to 
invert Eq.~(\ref{eq:Dinverse}) as in the symmetric case.

\section{A simple application}

The simplest possible dirty mesoscopic N--S 
system arguably consists of a disordered cavity or
junction with just one normal and one superconducting
lead.  In such a system the phase of the superconducting
order parameter can be gauged away, and so we assume that
$\phi=0$ and $f$ is real.  We denote the relative width of
the superconducting lead by $n_s = W_1/N$, so that the
conductance we seek is a simple function of
$0 < n_s < 1$, multiplied by the appropriate units and the
total number of modes $N$.

From the Schwinger-Dyson equation, Eq.~(\ref{eq:sd1}), we see
that
\begin{equation}
g  \> = \>  {1\over {1+f}}  \; ,
\label{eq:g}
\end{equation}
or equivalently
\begin{equation}
\alpha  \> = \>  {n_s \over {1+f}}  \; ,
\label{eq:sd2}
\end{equation}
 from Eq.~(\ref{eq:gamma}).  Putting Eqs.~(\ref{eq:alpha}) and 
(\ref{eq:g}) together, we obtain
\begin{equation}
{f \over {1+f^2} }  \> = \>  {n_s \over {1+f}}  \; ,
\label{eq:quad}
\end{equation}
and thus we determine $f$ completely in terms of $n_s$:
\begin{equation}
f  \> = \>  { {\sqrt{1+4n_s-4n_s^2}-1}\over {2(1-n_s)} }  \; .
\label{eq:formulaforf}
\end{equation}
The plus root is chosen so that as the number of modes
$n_s$ in the lead connecting the cavity and the
superconductor goes to zero, $f$ should vanish as is
physically reasonable.  The amplitude $f$ grows with
$n_s$, and approaches 1 when $n_s$ approaches 1.
In particular, for the symmetric case,
$n_s={1\over 2}$ we have $f={\sqrt{2}}-1$.

We proceed to evaluate the diffuson for this system.
Using Eq.~(\ref{eq:Dinverse}) with
\begin{equation} \label{tforgbp}
t  \> = \>  n_s g^2 \tau  \> = \>  {f \over {(1+f^2)(1+f)} } \tau  \; ,
\end{equation}
we find through simple arithmetic that
\begin{equation}
D  \> = \>  {{(1+f^2)(1+f)^2}\over(1+2f-f^2)}
    \left( I + {f (1-f)\over (1+f)} \tau 
\vphantom{\mbox{$1\over 2$}} \right) \; .
\label{eq:finalD}
\end{equation}

Now we are ready to compute the conductance, as
represented diagrammatically in Fig.~5(f).
As shown in the figure, there are two distinct terms which
contribute to this quantity --- one in which the
electrons are Andreev reflected ``individually'' (which
involves only $\Sigma$), and one in which the
top line and the bottom line in the diagram are connected,
which involves the diffuson $D$.

In the case of the non--diffuson contribution, one must
note that the appropriate terms of $G$ must contain only
a single copy of $\Sigma$, as opposed to the infinite
series reflected in Eq.~(\ref{eq:g}).  This is due to the fact
that the external indices represent modes in the normal
lead, and cannot be directly Andreev reflected ($G_0$ for
the normal lead vanishes --- excitations leaving the
cavity through it never return).
Thus, the non--diffuson contribution to the conductance is
given by
\begin{equation}
\sigma_\Sigma  \> = \>  4 {e^2 \over h} N (1-n_s) f^2  
     \> = \>  4 {e^2 \over h} N(n_s -f)  \; .
\label{eq:non}
\end{equation}
This is plotted in Fig.~6 as a function of $n_s$.

The conductance due to the diffuson is given by
\begin{equation}
\sigma_D  \> = \>  4 {e^2 \over h} N (1-n_s)^2 D_c  \> = \>
4 {e^2 \over h} N(1-n_s)^2 {f(1-f^4)\over 1+2f-f^2}  \; ,
\label{eq:diffus}
\end{equation}
which is also plotted in Fig.~6.
Note that to obtain the non--diffuson contribution to the
conductance we multiply by the number of modes on the
external lead $W_2 = (1-n_s) N$, while to obtain the
diffuson contribution we multiply by the square of this
number. In the non--diffuson contribution to the
conductance, an electron injected into a given mode comes
back as a hole in the same mode: this is known as the
``giant backscattering peak'' \cite{Been_GBP}.
In contrast, in the diffuson contribution, the hole
comes back in general in some other mode.

The total conductance is of course given by the sum of 
Eqs.~(\ref{eq:non}) and (\ref{eq:diffus}).  We find that 
when we add these two contributions to the conductance,
we obtain the surprisingly simple expression
\begin{equation}
\sigma_{NS} =  \sigma_\Sigma + \sigma_D = 
 2 {e^2 \over h} N \left( 1-{1\over \sqrt{1+4n_s-4n_s^2}} 
\vphantom{\mbox{$1\over 2$}} \right)
\; .
\label{eq:total}
\end{equation}
Indeed, if we introduce the asymmetry parameter $a$ by
$n_s=(1+a)/2$, so that $a=0$ for a symmetric cavity, 
we obtain simply
\begin{equation}
\sigma_{NS} = 2 {e^2 \over h} N 
      \left( 1-{1\over \sqrt{2-a^2}} 
\vphantom{\mbox{$1\over 2$}} \right)  \; .
\label{eq:total'}
\end{equation}
This is again plotted in Fig.~6.  For the case of equal widths 
of the leads, $a=0$, this has been derived using the
transmission--eigenvalue approach in Ref.~\cite{JPB}. 
Incidentally, note that $\sigma_{NS}$ is symmetric with respect 
to interchanging the widths of the leads, $a \leftrightarrow -a$. 
In the transmission--eigenvalue approach, this follows directly 
from the fact that $\sigma_{NS}$ depends only on the transmission 
eigenvalues \cite{Been1} (even when these are appropriately 
redefined for rectangular transmission matrices).

For comparison, we consider the case in which the charge
conjugation symmetry between electrons and holes is broken,
by a finite bias voltage $V$, a finite temperature, or a
finite magnetic field.  As explained above
(see the end of Sec.~II~A and Sec.~II~C) we may then 
denote the scattering matrix of holes by a unitary 
matrix $S'$, which to leading order can be taken as 
unrelated to $S$, and average over $S$ and $S'$ separately.
The situation becomes enormously simpler: all the
non--diffuson graphs cease to exist, $\Sigma=0$, and we 
need only to evaluate the diffuson contribution with 
$\Gamma=I$ and $G=G_0$.  We are left with the ladder graphs
shown in Fig.~7, with the number of rungs on the ladder
restricted to be even.  The conductance is thus given by
$(1-n_s)^2(n_s+n_s^3+n_s^5+...)=
n_s({{1-n_s}\over{1+n_s}})={{(1-a)(1+a)}\over{2(3+a)}}$,
which we plot in Fig.~8. Notice that it is not symmetric
in the asymmetry parameter $a$.  We also plot the
difference between the total conductance with and without
charge conjugation (or time reversal) symmetry. We see that
for $n_s < {1\over 2} (1+a_c) \sim 0.65$ charge conjugation 
symmetry actually lowers the conductance 
($a_c = 0.2955 \dots$ is the real solution of 
$a_c^3+a_c^2+3a_c-1=0$).

Incidentally, the total conductance without charge
conjugation symmetry can be calculated easily using
elementary physics. Since the scattering of the hole in
the cavity is no longer strongly correlated with the
scattering of the electron, the system is analogous to
two independent cavities connected in series, the motion
of electrons being represented as excitations in the
first cavity, and holes being represented as excitations
in the second.  The conductance of such a two--cavity
system is given, to leading order, simply by the series
addition of the resistances corresponding to the leads
that connect them to each other and to the
``electron electrode'' and its copy, the
``hole electrode'':
$\sigma_{2C} =
2 {e^2 \over h} (W_2^{-1} + W_1^{-1} + W_2^{-1})^{-1}$
(the subscript $2C$ implies 2 cavities).  According to
Eq.~(\ref{eq:conductance}), the conductance of the N--S structure
is given by twice this value, which is, as noted above,
\begin{equation} \label{nonsymm}
\sigma_{NS}^{\rm non-symmetric}  \> = \>
4N {e^2 \over h} n_s {1-n_s \over 1+n_s}  \; .
\end{equation}
Another comparison may be made with the conductance 
the same structure would have if both electrodes were 
normal, $\sigma_N = (2e^2/h) \, N \, n_s (1 \! - \! n_s)$.
However, experimentally it is usually much easier to 
observe the crossover from $\sigma_{NS}$ to 
$\sigma_{NS}^{\rm non-symmetric}$ than that from
$\sigma_{NS}$ to $\sigma_N$.

The results reported here are in complete agreement with
previous theoretical work.  For example, the fact that
breaking electron--hole symmetry may either decrease or 
increase the conductance, $\sigma_{NS}$, is known, 
at least in principle \cite{explain}.  This 
situation suggests the following experimental challenge: 
to fabricate a device with an ergodic scatterer connected 
to electrodes through leads with tunable widths, for 
example by changing a gate voltage in a two--dimensional 
electron gas device, such that $\sigma_{NS}$ would exhibit 
positive magnetoconductance for some gate voltages and
negative magnetoconductance for other gate voltages.  The
challenge here is considerable because fabricating good
contacts between a two--dimensional electron gas and a
superconducting electrode is not easy --- there are
invariably large mismatches in the Fermi velocity, not to 
mention Schottky barriers, etc.
In fact, the goal of fabricating such contacts has been 
one of the driving forces behind the development of this 
field in the past few years.  Note that related 
normal--normal conductances have already been shown 
to exhibit changes in the sign of the magnetoconductance 
as a function of geometric parameters of the system, in 
Ref.~\cite{Petr_HM}. 

It is interesting to compare our results with the 
experimental data of Kastalsky {\it et al.\ }\cite{Kast}, 
who have studied an N--S system in which a relatively large 
piece of ``normal metal'' (in this case, a degenerate 
semiconductor) was in contact with a superconductor and a 
normal reservoir (the semiconducting substrate).  The
conductance of this junction $\sigma_{NS}$ was observed
to grow by a factor of about 1.8 upon decreasing an applied 
magnetic field at low temperatures (the temperature dependence 
is complicated by not having a very large separation between 
the Thouless energy and the superconducting gap).  This 
compares favorably with the results shown in Fig.~8 for 
large $n_s$ --- in this case $f$ approaches 1 while the 
resistance of the wide lead, $W_1$, becomes negligible; 
the conductance due to ``direct'' Andreev reflections 
($\sigma_\Sigma$) is then equal to twice that which is 
obtained when the holes must ``find their way to the external 
electrodes'' independently of the motion of the electrons 
that produced them [expanding Eqs.~(\ref{eq:total}) and 
(\ref{nonsymm}) near $n_s=1$ gives 
$\sigma_{NS} \rightarrow {4e^2 \over h} W_2$ and
$\sigma_{NS}^{\rm non-symmetric} \rightarrow {2e^2 \over h} W_2$
respectively].   This experiment was in fact one of the first
available in the subfield of mesoscopic N--S structures
discussed here, and was initially very hard to interpret.
The interpretation which was eventually accepted
\cite{vanWees} put forward precisely the ideas of multiple
Andreev reflection which are formalized in the present
work and its predecessors.

\section{Conclusions and outlook}

We have developed a new technique for evaluating the
dissipative conductances of disordered mesoscopic 
systems attached to normal and to superconducting leads.  
It is based on a generalization of the 
planar--diagrammatic technique of large--$N$ random 
matrix models, which allows one to consider unitary 
(as opposed to Hermitean) matrices, specifically the 
scattering matrix of a disordered mesoscopic grain.  
The leading order results for the conductance are 
affected by mesoscopic coherence, because of the 
possibility and importance of an exact symmetry 
between electrons and holes at the Fermi level.

In the present work, we have demonstrated the method 
only by application to the simplest possible disordered 
mesoscopic N--S system: a scattering cavity attached 
through ideal leads to one superconducting electrode 
and one normal electrode.  This calculation can only 
serve as a ``toy--model'' description of actual 
experimental geometries; it is necessary to consider 
many possible complications, such as the effects of 
potential barriers, in order to check, e.g., whether 
there is any physics behind the agreement with the 
experiment of Ref.~\cite{Kast} which was mentioned at 
the end of the previous section.  This underlines the 
importance of generalizing the present method to deal 
with more complicated geometries (in analogy with the 
``circuit theory of Andreev conductance'' put forward 
by Nazarov \cite{Nazarov}).  Such a generalization turns 
out to be possible, and the results will be reported 
in future work \cite{AZ}.

An additional possible direction for future pursuit is
to evaluate the fluctuations and the next order
corrections to the conductances.  Specifically, the
much--discussed $O(N^0)$ term can in principle be
found from similar large--$N$ diagrams which break
the condition of planarity exactly once.  A third
direction is to consider excitations with finite
energies, or systems in a finite magnetic field.
Such calculations could turn out to be much more difficult, 
because the scattering matrices depend on energy and magnetic
field in a continuous manner, and it should be necessary
to invoke ensembles of such two--parameter families of
scattering matrices.  However, the experience in
mesoscopic phenomena may lead to a hope that the generic
behavior in such ensembles would be simple and universal
\cite{Altsh}, perhaps describable by simple modifications of
the $c_k$ couplings used in the present method.

\section*{Appendix}
%\bskip24pt

Here we give an alternative derivation of (\ref{eq:cks}). As customary in
the literature on random matrix theory, we define the function of a complex
variable $z$ called the ``one--point'' Green's function by
\begin{equation}
G(z) \> = \>  \left\langle  {1\over N} {\rm tr} {1 \over z-(S+S^\dagger)} 
\right\rangle
\label{eq:A1}
\end{equation}
(which should not be confused with the quantity $G$ in the main text). 
As before, $\left\langle \dots \right\rangle$ denotes averaging the 
$N$ by $N$ unitary matrix $S$ over the CUE.
Diagonalizing the unitary matrix $S$ and denoting its eigenvalues by
$e^{i\theta_j}$, $j=1,2,....N$, we may use the obvious fact that the 
eigenvalues are distributed uniformly over the unit circle to find that
\begin{equation}
G(z)  \> = \>  \int {d\theta\over 2\pi} \, {1\over z-2 cos\theta}
  \> = \>  \sum_{k=0}^{\infty}{1 \over z^{2k+1}}{(2k)!\over (k!)^2}
  \> = \>  {1\over \sqrt{z^2-4}}  \; ,
\label{eq:A2}
\end{equation}
where the sum converges for $|z| > 2$.

On the other hand, if we evaluate (\ref{eq:A1}) by a
diagrammatic expansion as shown in Fig.~9, we have the
two equations
\begin{equation}
G(z)  \> = \>  {1\over z-\Sigma(z)}
\label{eq:defini}
\end{equation}
 and
\begin{equation}
\Sigma(z)  \> = \>  2 \sum_{k=1}^{\infty}\gamma_{2k}G^{2k-1}(z) 
\label{eq:sig}
\end{equation}
(the latter holds at least to leading order in $1/N$).  
We recognize (\ref{eq:defini}) as just the Schwinger--Dyson equation 
again: here the bare Green's function $G_0(z)$ is simply equal to $1/z$.
In (\ref{eq:sig}) we have used the definition of $\gamma_{2k}$
as the coupling constant involving $2k$ matrix elements. Note
also the factor of $2$: the first matrix element in the diagram for
$\Sigma$, reading from left to right say, may represent either $S$ 
or $S^\dagger$. 
This feature does not appear in discussions of Hermitean random matrices.

Our goal here is to determine the coupling constants $\gamma_{2k}$,
assuming that averages over elements of $S$ and $S^\dagger$ may indeed
be written in terms of such diagrams, with a generalization of Wick's
theorem implying a multiplicative property of the weights of diagrams
involving more than one coupling (those of Fig.~3). We do
this by eliminating $z$ between (\ref{eq:A2}) and (\ref{eq:defini}) and
thus solving for $\Sigma(z)$ in terms of $G(z)$:
\begin{equation}
\Sigma(z)  \> = \>  {\sqrt{1+4G^2(z)}-1\over G(z)}  \; ,
\label{eq:A5}
\end{equation}
which is identical to Eq.~(\ref{eq:w}) if we identify $x = -G^2(z)$ and
$w = - \Sigma(z) G(z) / 2$ [cf.\ also Eq.~(\ref{eq:alpha})].
Expanding the right-hand side as a series in $G$ and comparing with
(\ref{eq:sig}) we obtain immediately $\gamma_{2k}$ and
hence $c_k$ in agreement with (\ref{eq:cks}).

\acknowledgments
%\section*{Acknowledgments}

The authors would like to thank J.~Feinberg for helpful 
discussions.
This work is supported in part by the National
Science Foundation under Grants No. PHY94-07194 and 
DMR93-08011.

\figure{ Fig.~1:
Sketch of a simple normal--superconducting mesoscopic
system, consisting of a single ``grain'' which is treated
as an ideally ergodic scattering cavity, connected to 
a normal electrode (particle reservoir) and to a 
superconducting electrode through ideal leads of 
different widths.
}

\figure{ Fig.~2:
An example of a diagram contributing to the unaveraged
conductance of Eq.~(\ref{eq:conductance}).
}

\figure{ Fig.~3:
The different couplings, and their weights [see 
Eq.~(\ref{eq:cks}) below].  These couplings are used to connect 
the double lines of Fig.~2.
}

\figure{ Fig.~4:
Diagrams corresponding to (a)
$\left\langle   S_{ia} S^\dagger_{ai} \right\rangle$,
(b) $\left\langle  S_{ia} S^\dagger_{aj} S_{jb} S^\dagger_{bi} 
\right\rangle$, and
(c) $\left\langle  S_{ia} S^\dagger_{aj} S_{jb} S^\dagger_{bk}
S_{kc} S^\dagger_{ci}  \right\rangle$.
}

\figure{ Fig.~5:
Diagrammatic representation of: (a) the ``self--energy''
$\Sigma$, (b) the ``bare Green's function'' $G_0$,
(c) the ``averaged Green's function'' $G$,
(d) the ``effective coupling constant'' $\Gamma$,
(e) the ``diffuson'' $D$, and (f) the conductance $\sigma$ 
of Eq.~(\ref{eq:conductance}).
}

\figure{ Fig.~6:
The conductance of the simplest normal--superconducting
system considered, in units of $4 N e^2/h$, as a function of
the fraction of modes belonging to the superconducting lead
(full line).  The non--diffuson (dashed) and the diffuson
(dotted) contributions are also shown separately, the former
corresponding to ``direct'' Andreev reflections from the
cavity, or the so--called giant backscattering peak.
}

\figure{ Fig.~7:
The simplified diagrams contributing to the conductance when
the symmetry between electrons and holes is broken.  Only 
vertical simple double lines can be used, because the diagram must 
be planar and the matrices on the even rungs of the 
ladder cannot be connected to the matrices on the odd rungs. 
}

\figure{ Fig.~8:
A comparison of the conductance of Fig.~6 (full line) to that
obtained in the absence of electron--hole symmetry (dashed
line).  Also shown is the difference between the two, on an
expanded scale (top panel).
}

\figure{ Fig.~9:
Diagrams for the one--point Green's function, $G(z)$ (thick 
lines), and its associated self--energy $\Sigma(z)$ (shaded 
semicircles) of the appendix.  The thin lines represent 
$G_0(z)=1/z$, and carry no arrows.  The double lines 
represent an element of $S$ if the arrow is pointing away 
from the single line, and an element of $S^\dagger$ in 
the opposite case.  Although the quantities $G(z)$ and 
$\Sigma(z)$ are not to be confused with $G$ and $\Sigma$ of 
the main text, the couplings are the same 
(see Fig.~3).
}


\begin{references}
\bibitem{corB} Again, one may estimate the correlation field
from the semiclassical analysis: the field required to shift
the phases of the orbits threads approximately one flux
quantum through the directed area encircled by a typical
orbit.  This area is of order $\sqrt{t_{esc}/t_{erg}}$ times
the area of the cavity, where $t_{erg}$ is the time required
for a particle to reach from one side of the cavity to the
other (the assumption of ergodic scattering implies 
$t_{esc} \gg t_{erg}$).
\bibitem{notation} Note that this notation is convenient only
for the single--cavity case; in Ref.~\cite{AZ} we use a
``bare Green's function'' which represents only the motion
through the leads and is diagonal in the electron--hole
subspace (the Andreev reflections are then included in a
suitable redefinition of $\Sigma$).
\bibitem{explain} The order--$N$  effects considered here do 
not occur in diffusive wires with large enough $L/l$.  Indeed, 
the emphasis on diffusive wires and tunnel contacts, e.g., in 
Refs.~\cite{Nazarov} and \cite{p}, may lead to the impression 
that electron--hole symmetry always enhances the conductance 
of the structure, except for the weak localization effect,
which decreases the conductance but is of higher order in 
$1/N$.  In fact, Ref.~\cite{Brou} went so far as to suggest 
the possibility of restricting the diffusive metallic regime 
to a range of parameters such that the weak localization 
correction is greater than the order--$N$ effect (which is 
referred to there as a magnetic--field-- and voltage--dependent 
``contact resistance'').  From the present point of view, 
based on notions of ideal leads and the large--$N$ limit, such 
a strong emphasis on weak localization is not warranted.
\bibitem{rmkby} Remarkably, corrections to $\Sigma(z)$ from 
higher order diagrams and from corrections to the products of 
coefficients $\gamma_{2k}$ (i.e., corrections to the $C_{u,v}$ 
coefficients), seem to cancel each other.  This follows from 
the fact that Eqs.~(\ref{eq:A2}) and (\ref{eq:defini}) hold 
for any $N$.
\end{references}
\end{document}